\documentclass[twocolumn,showpacs,preprintnumbers,amsmath,amssymb]{revtex4-1}
\usepackage{graphicx}
\usepackage{dcolumn}
\usepackage{bm}
\usepackage{color}

\begin{document}
\title{A novel interpretation of the ``$\Theta^{+}(1540)$ pentaquark" peak}
\author{A. Mart\'inez Torres}
\author{E.~Oset}
\affiliation{ Departamento de F\' isica Te\' orica and IFIC, Centro Mixto Universidad de Valencia-CSIC, Institutos de Investigaci\'on
de Paterna, Aptd. 22085, 46071 Valencia, Spain.}
\date{\today}
\begin{abstract}
 We use a theoretical model of the $\gamma ~d \to ~K^+ K^- ~n ~p $ reaction adapted to the experiment done at LEPS where a peak was observed and associated to the $\Theta^{+}(1540)$ pentaquark. The study shows that the method used in the experiment to associate momenta to the undetected proton and neutron, together with the chosen cuts, necessarily creates an artificial broad peak in the assumed $K^+ n$ invariant mass in the region of the claimed $\Theta^{+}(1540)$, such that the remaining strength seen for the experimental peak  is compatible with a fluctuation of 2$\sigma$ significance.
 \end{abstract}

\maketitle

In the work of  \cite{nakaone}
 the $\gamma\,\, ^{12}C \to K^+ K^- X$ reaction was studied and a peak was found in the $K^+n$ invariant mass spectrum around 1540 MeV, which was identified as a signal for a pentaquark of positive strangeness, the ``$\Theta^+$''. The unexpected finding lead to a large number of poor statistics experiments where a positive signal was also found, but gradually an equally big number of large statistics experiments showed no evidence for such a peak. 
A comprehensive review of these developments was done in \cite{hicks}, where one can see the relevant literature on the subject, as well as in the devoted section of the PDG \cite{pdg}. 

 More recently a new experiment was done at LEPS on a deuteron target, 
and with more statistics, and a clear peak was observed around 1526 MeV in the
$K^+n$ invariant mass distribution \cite{nakatwo}. Yet, the experiment of  \cite{kinnon} dealing with the same reaction as in LEPS but with ten times more statistics and with complete kinematics (but excluding small angles), failed to see any peak around the ``$\Theta^+$" region. The detail of complete kinematics should be stressed because in the LEPS experiment neither the proton nor the neutron were measured and an educated guess had to be made for their momenta. 

In order to understand what is behind the peak seen in \cite{nakatwo}, we have constructed a theoretical model which contains the basic ingredients seen in the LEPS experiment, $\phi$ production on the proton and the neutron and $\Lambda(1520)$ production on the proton, together with rescattering of the kaons. The details of this model can be seen in \cite{pentafirst}. We adapt the model to the set up of the LEPS experiment, generating twenty random energies between  2 GeV to 2.4 GeV and implement the angular cuts and the mass cuts to eliminate the $\phi$ peak. 

The LEPS detector is a forward magnetic spectrometer. Its geometry is implemented in our simulation by imposing that the angle of the kaons in the final state with respect the incident photon is not bigger than 20 degrees. 

The nucleons are not detected at LEPS, therefore, some prescription is required in order to estimate the momentum of the $p$ and $n$ in the reaction  $\gamma d \to K^{+}K^{-}np$ and determine the invariant mass of $K^{-}p$ or $K^+ n$. This is done
using the minimum momentum spectator approximation (MMSA). For this purpose one defines the magnitude 
\begin{align}
p_{pn}=p_{miss}=p_{\gamma} +p_d - p_{K^+} -p_{K^-}
\end{align}
which corresponds to the four momentum of the outgoing $pn$ pair. From there one evaluates the nucleon momentum in the frame of reference where the $pn$ system is at rest, $\vec{p}_{CM}$.
Boosting back this momentum to the laboratory frame, we will have a minimum modulus for the momentum of the spectator nucleon when the momentum $\vec{p}_{CM}$ for this nucleon goes in the direction opposite to $\vec{p}_{miss}$. Thus, the minimum momentum, $p_{min}$, is given by 
 \begin{align}
 p_{min} = -|\vec{p}_{CM}| \cdot \frac{E_{miss}}{M_{pn}} + E_{CM} \cdot\frac{|\vec{p}_{miss}|}{M_{pn}}
\end{align}
where $E_{CM}=\sqrt{|\vec{p}_{CM}|^{2}+M^{2}_{N}}$ is the energy of the nucleon in the CM frame. In this case, the momentum of the other nucleon will be in the direction of the missing momentum  with a magnitude 
\begin{align}
p_{res} = |\vec{p}_{miss}| -p_{min}	
\end{align}
In \cite{nakatwo} the $M_{K^+ n}$ invariant mass for the reaction $\gamma d\to K^{+}K^{-}np$ is evaluated assuming the proton to have a momentum $p_{min}$ (actually what one is evaluating is $M_{K^+ N}$, with $N$ the non spectator nucleon). Consequently, in this prescription, the momentum of the neutron in the final state will be
\begin{align}
\vec{p}_{n}=p_{res}\cdot\frac{\vec{p}_{miss}}{|\vec{p}_{miss}|}
\end{align}
which is used to calculate the $M_{K^{+}n}$ invariant mass for the reaction $\gamma d\to K^{+}K^{-}np$ in \cite{nakatwo}.  A cut is imposed at LEPS demanding that  $|p_{min}|<$ 100 MeV. This condition is also implemented in our simulation of the process.

In order to remove the contribution from the $\phi$ production at LEPS one considers events which satisfy that the invariant mass of the $K^{+}K^{-}$ pair is bigger than 1030 MeV and bigger than the value obtained from the following expression
\begin{align}
1020\, \textrm{MeV} +0.09\times (E^{eff}_{\gamma}(\textrm{MeV})-2000\, \textrm{MeV})
\end{align}
where $E^{eff}_{\gamma}$ is defined as the effective photon energy
\begin{align}
E^{eff}_{\gamma}=\frac{s_{K^{+}K^{-}n}-M^{2}_{n}}{2M_{n}}\label{Eeff}
\end{align}
with $s_{K^{+}K^{-}n}$ the square of the total center of mass energy for the $K^{+}K^{-}n$ system calculated using the MMSA approximation to determine the momentum of the neutron assuming the proton as spectator. In \cite{nakatwo} only events for which $2000$ MeV $<E^{eff}_{\gamma} $ $<2500$ MeV are considered, a condition which is also incorporated in our simulation. The $E^{eff}_{\gamma}$ of Eq. (\ref{Eeff}) with the MMSA prescription would correspond to the photon energy in the frame where the original non spectator (participant) nucleon is at rest.

\begin{figure}[h!]
\centering
\includegraphics[width=0.35\textwidth]{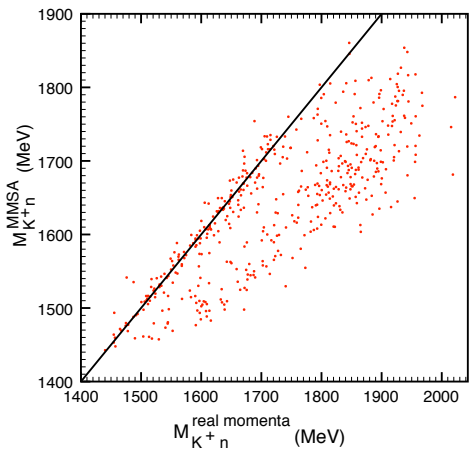}
\caption{$M_{K^{+}n}$ calculated using the MMSA prescription versus $M_{K^{+}n}$ obtained with the real momentum for the nucleons and the full model, i.e., $\phi$ production on the nucleons and $\Lambda(1520)$ production on the proton.}\label{Minvnaka_real}
\end{figure}

The first result that we show is the distribution of $K^{+}n$ invariant masses for the LEPS set up using the real momenta obtained from our Monte Carlo integral of the cross section versus the one obtained using the momenta determined with the MMSA prescription (Fig. \ref{Minvnaka_real}). We see two blocks of points, one of them sticking  around the diagonal and another one with points scattered around the plane. This is so, because actually the MMSA prescription reconstructs the $K^+ N$ invariant mass, where $N$ is the participant nucleon, which about half of the times is the neutron and the other half the proton. The points around the diagonal correspond to the case where the participant is the neutron.

\begin{figure}[h!]
\centering
\includegraphics[width=0.35\textwidth]{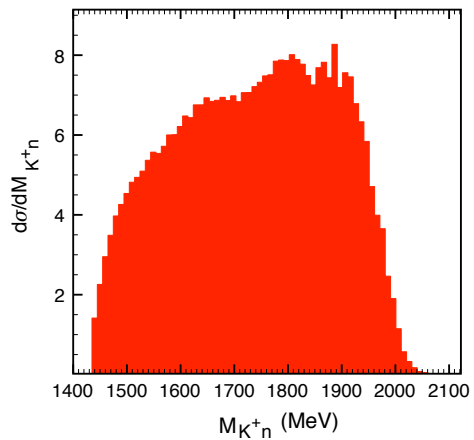}
\caption{$M_{K^{+}n}$ invariant mass distribution calculated using the real momenta and with a $\phi$ cut of $M_{K^{+}K^{-}}>1050$ MeV.}\label{phi1050}
\end{figure}

\begin{figure}[h!]
\centering
\includegraphics[width=0.35\textwidth]{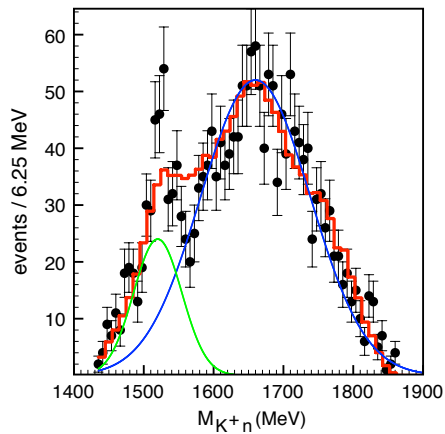}
\caption{$M_{K^{+}n}$ invariant mass distribution obtained with the MMSA prescription and same cuts than LEPS normalized to the data of \cite{nakatwo} (shown as dots), together with two gaussians functions: one peaking at 1520 MeV with a width of 80 MeV and another one peaking at 1660 MeV with a width of 185 MeV.}\label{Kplusnnorm1}
\end{figure}

The association of the $K^+ N$ spectrum to $K^+ n$ at LEPS has a repercussion in the assumed experimental $K^+ n$ distribution. The real distribution from the model is given in Fig. \ref{phi1050}.
We have to take a cut for the $K^+K^-$ invariant mass slightly different than in \cite{nakatwo} because the cut at LEPS involves $E_{\gamma}^{eff}$ which relies for its definition on the MMSA prescription itself, so we must avoid that to use the real momenta.  As we can see in Fig. \ref{phi1050}, there is no peak of the distribution around the ``$\Theta^+$" peak. Instead we show in Fig. \ref{Kplusnnorm1} the distribution obtained using the LEPS cuts and the MMSA prescription, normalized to the experimental data, which are also shown in the figure. 
 
We use the full model of \cite{pentafirst} including $\phi$ production on the proton and the neutron plus $\Lambda(1520)$ production on the proton.  The MMSA prescription is used to compare to the data as explained before. As we can see in Fig. \ref{Kplusnnorm1}, the combination of the LEPS cuts and the MMSA prescription, which also affects the cut, has produced an artificial peak below the region of the  ``$\Theta^+$". We show in the figure how the shape of the distribution can be represented in terms of two gaussians, one of them peaking around the ``$\Theta^{+}$" peak. This means that in a large statistics experiment one would see this clear broad peak, which could be interpreted as a sign of a resonance. Yet, there is no resonance in that region in the model used.  Coming back to the comparison of the ``exact" distribution, generated from the theoretical model, with the data (taken from Fig. 12 a) from \cite{nakatwo}), we see that the ``$\Theta^+$" peak has three points on top of the ``exact" distribution. The accumulated strength of these three bins over the ``exact" curve is about 35 events.  The question now remains: could this peak, measured from the ``exact" curve, be a statistical fluctuation due to the limited number of events of the experiment (around 2000 events)?.  A hint to  answer this question is provided by the LEPS experiment in fig. 10 a) of \cite{nakatwo}, where a peak followed by a dip is seen on top of the assumed background in the $K^-p$ mass distribution. In order to compare with these data, 
\begin{figure}[h!]
\centering
\includegraphics[width=0.35\textwidth]{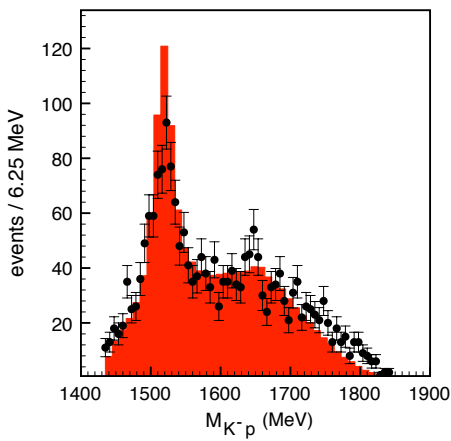}
\caption{$M_{K^{-}p}$ invariant mass distribution calculated with the MMSA prescription and same cuts than LEPS normalized to the data of \cite{nakatwo} (shown as dots).}\label{Kmpnorm1}
\end{figure}
we show in Fig. \ref{Kmpnorm1} the mass distribution for $K^-p$ obtained with our model and the MMSA prescription (actually the $K^- N$ distribution, as discussed above), together with the data (Fig. 10 a) of \cite{nakatwo}). The agreement with experiment is good, like for the $K^{+}n$ distribution shown in Fig. \ref{Kplusnnorm1}, indicating that we have indeed a realistic model. The discrepancies in the large momentum tail can be cured adding a small contribution of the broad ($\Gamma$= 300 MeV) $\Lambda$(1800) production to account for the tail of the $K^-$ p distribution, which also reduces a bit the peak of the $\Lambda(1520)$ upon normalization to the data, and we found minor changes in the $K^+$ n distribution, less than 5 $\%$ in the region of the ``$\Theta^+$'' peak. Yet, the point we want to make is that there is a peak in the data around 1650 MeV with four points over the ``exact" curve with an accumulated strength of about 33 events, followed by a similar dip, a typical structure of a statistical fluctuation.  Since for this distribution there are no discrepancies in the interpretation with respect to \cite{nakatwo} we can take the background from there (solid line of Fig. 10 a) of \cite{nakatwo}) and we also find about 40 events. This peak was not associated to any resonance in \cite{nakatwo}. Instead it was dismissed as a statistical fluctuation. In fact, the peak disappeared when a slightly different cut was made (see Fig. 16 (Right) of \cite{nakatwo}).
The interpretation of this peak and dip as a fluctuation, as done in \cite{nakatwo}, is fair, as this peak has about $2 \sigma$ significance over the background ($\sigma$ is the statistical error of the data for which $\sqrt {N_{events}}$ is taken in \cite{nakatwo}). But the peak of the ``$\Theta ^+$" over the ``exact" distribution has also about 35 accumulated events and consequently could also be considered as a statistical fluctuation. 

 Here we would like to make a more quantitative study of the statistical significance of the ``$\Theta^+$" peak. In \cite{nakatwo} a best fit to the data was done assuming a background and a Gaussian peak in the  ``$\Theta^+$" region. The best fit with these assumptions provided a background of about 22 events per bin below the ``$\Theta^+$" peak, as can be seen in Fig. 12 a) of \cite{nakatwo}. With respect to this background, the ``$\Theta^+$" peak has a strength of about $5 \sigma$. According to this, the statistical significance of the peak would rule out the possibility of it being a statistical fluctuation. 
 
Conversely, after the theoretical evaluation of the background, our argumentation goes as follows: The actual background below the ``$\Theta^+$" peak is bigger than the one provided by the LEPS best fit, around 36 events per bin instead of the 22 assumed in \cite{nakatwo}. This makes the strength of the peak with respect to the background much smaller than in the LEPS best fit. It also makes $\sigma$ larger and the statistical significance is now of about $2 \sigma$, something acceptable as a fluctuation, as in the case of the experimental $K^- p$ spectrum. This argumentation about the significance of the peaks is corroborated by further calculations which we have carried out. First, the best fit of LEPS is not the only good fit possible. We have seen that a fit to the data with a background and a fluctuation (a peak followed by a dip) gives the same reduced $\chi^2$ than in \cite{nakatwo}, but returns a background nearly identical to the calculated one. Second, in order to know the actual errors of the limited LEPS statistical we have made 10 runs with the Von Neumann rejection method, producing about 2000 events each, like in \cite{nakatwo}. This method proceeds like the experiment, generating events or not according to their probability to be produced, and we have checked that the statistical significance of the runs is equivalent to that of the experiment. From these runs we have evaluated the statistical errors of each run (see section VI of \cite{pentafirst}). The errors found are of the order of 20\% in the region of the ``$\Theta^+$". This means an error ($\sigma$) of about 7 events per bin, such that the difference of the peak to the background is indeed of the order of $2 \sigma$.

The other point we want to make is that it is possible to produce peaks by changing the cuts.  This is shown explicitly in the experiment of \cite{nakatwo} since the mentioned  peak at 1650 MeV in the $K^-p$ mass distribution in Fig. 10 a) of that paper disappears when a slightly different cut is made in Fig. 16 (Right) of the same paper. This should be a warning for analyses in this kind of problems, which we want to make manifest by showing it also in our calculations.

\begin{figure}[h!]
\centering
\includegraphics[width=0.35\textwidth]{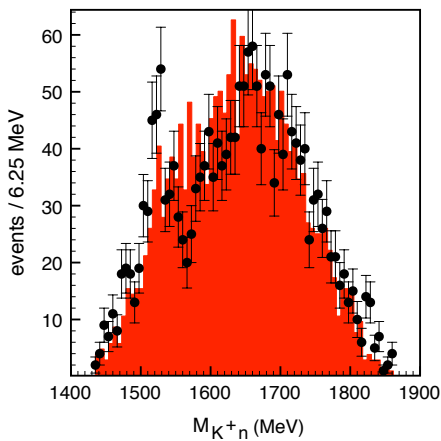}
\caption{$M_{K^{+}n}$ invariant mass distribution calculated with $\sim$ 2000 events, the MMSA prescription and the cut $M_{K^{+}K^{-}}>1030$ MeV compared with the data of \cite{nakatwo} (shown as dots).}\label{VN1}
\end{figure}

\begin{figure}[h!]
\centering
\includegraphics[width=0.35\textwidth]{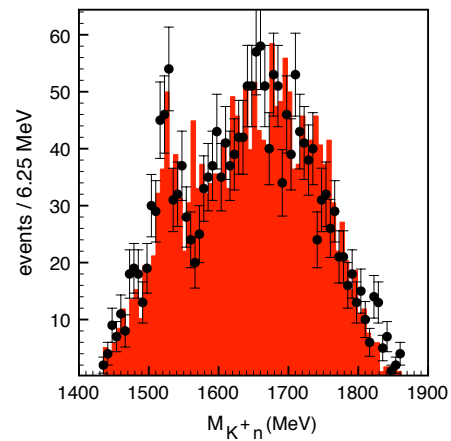}
\caption{$M_{K^{+}n}$ invariant mass distribution calculated with $\sim$ 2000 events, the MMSA prescription and the same cuts as those made in \cite{nakatwo} compared with the data of \cite{nakatwo} (shown as dots).}\label{VN2}
\end{figure}

For this purpose, in Fig. \ref{VN1} we show the results obtained for a chosen Von Neumann run, with about 2000 events, and the MMSA prescription with the cut $M_{K^+ K^-}>$ 1030 MeV. In Fig. \ref{VN2} we show our results for the same run when the cut is the one of the main LEPS set up discussed previously in the present paper. 

We can see that a peak is generated in the region of 1530 MeV. This is a chosen example to show what can happen with the use of cuts, but the fact is that in any case, for a given run with a certain number of events (equivalent to a given experiment), it is possible to get some enhancement in a chosen region by making small variations of the cuts. 

 In summary, our study has shown that the background in the $\gamma ~d \to ~K^+ K^- ~n ~p $ reaction is fairly larger than the one obtained in the best fit to the data of LEPS assuming a background and a Gaussian peak in the region of the  ``$\Theta^+$". We also mentioned that the fit of LEPS is not unique and other fits to the data, assuming a background and a fluctuation, are possible, producing the same reduced $\chi^2$ and returning a background nearly identical to the calculated one. Based on the calculated background and the errors obtained from different Monte Carlo Von Neumann runs, we evaluated the statistical significance of the ``$\Theta^+$" peak and found it to be of about $2 \sigma$ with respect to the background, compatible with a fluctuation. The larger statistical significance claimed in \cite{nakatwo} was tied to the assumption of a significantly smaller background, which we have found is not justified. \\

This work is partly supported by the DGICYT contract  FIS2006-03438,
the Generalitat Valenciana in the program Prometeo and 
the EU Integrated Infrastructure Initiative Hadron Physics
Project, n.227431. A. M. T thanks the support from a FPU grant
of the MICINN.

\end{document}